\documentclass[doublecol]{epl2} 

\usepackage{epsfig,amsmath,amssymb}
\usepackage{graphics} 
\usepackage{subfigure}
\usepackage{color}
\usepackage{multirow}
\usepackage{mathrsfs}

\title{Spin-gap study of the spin-$\frac{1}{2}$ $J_{1}$--$J_{2}$ model on the triangular lattice}
\shorttitle{Spin-$\frac{1}{2}$ $J_{1}$--$J_{2}$ model on the triangular lattice} 

\author{R. F. Bishop\inst{} \and P. H. Y. Li\inst{}}
\shortauthor{R. F. Bishop \etal}

\institute{                    
  \inst{}  School of Physics and Astronomy, Schuster Building, The University of Manchester, Manchester, M13 9PL, UK
}
\pacs{75.10.Jm}{Quantised spin models, including quantum spin frustration}
\pacs{75.10.Kt}{Quantum spin liquids, valence bond phases and related phenomena}

\abstract{ 
We use the coupled cluster method implemented at high
  orders of approximation to study the spin-$\frac{1}{2}$ $J_{1}$--$J_{2}$
  model on the triangular lattice with Heisenberg interactions between
  nearest-neighbour and next-nearest-neighbour pairs of spins, with
  coupling strengths $J_{1}>0$ and $J_{2} \equiv \kappa J_{1} >0$,
  respectively.  In the window $0 \leq \kappa \leq 1$ we find that
  the 3-sublattice 120$^{\circ}$ N\'{e}el-ordered
  and 2-sublattice 180$^{\circ}$ stripe-ordered antiferromagnetic
  states form the stable ground-state phases in the regions $\kappa <
  \kappa^{c}_{1} = 0.060(10)$ and $\kappa > \kappa^{c}_{2} =
  0.165(5)$, respectively.  The spin-triplet gap is found to vanish
  over essentially the entire region $\kappa^{c}_{1} < \kappa <
  \kappa^{c}_{2}$ of the intermediate phase.  }

\begin{document}

\maketitle

\begin{figure*}[!tbh]
\begin{center}
\mbox{
\subfigure[]{\includegraphics[height=2.7cm]{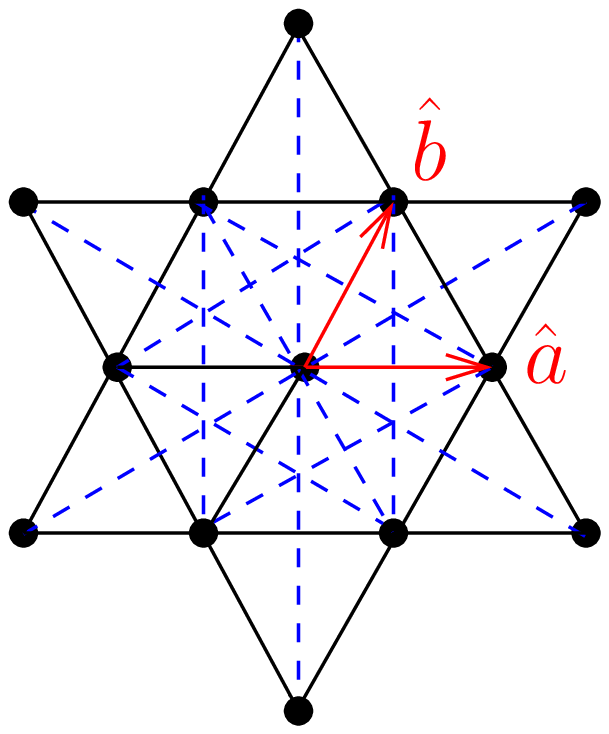}}
\quad
\subfigure[]{\includegraphics[height=2.7cm]{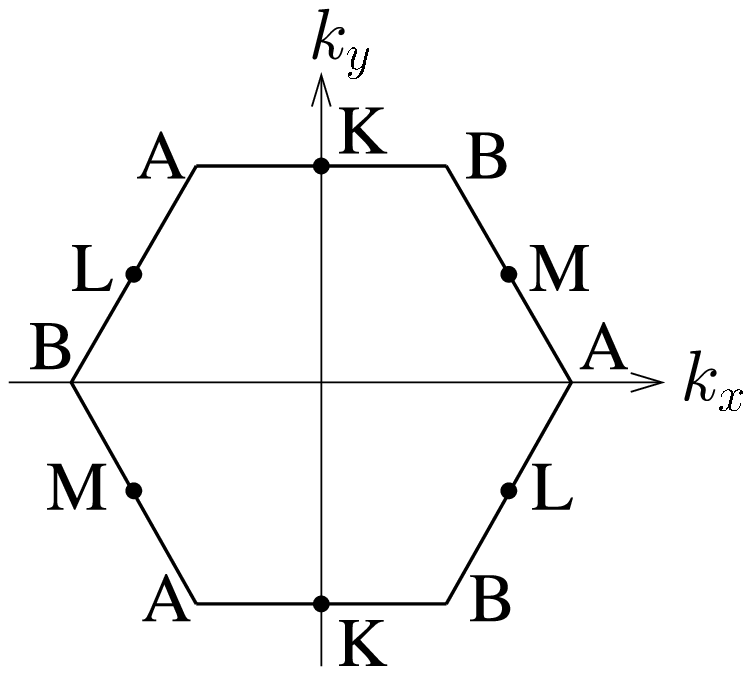}}
\quad
\subfigure[]{\includegraphics[height=2.7cm]{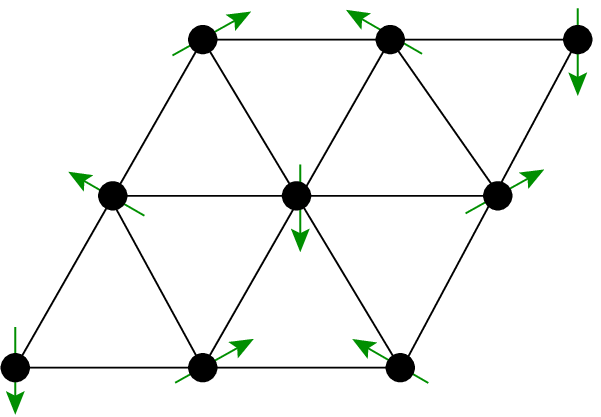}}
\quad
\subfigure[]{\includegraphics[height=2.7cm]{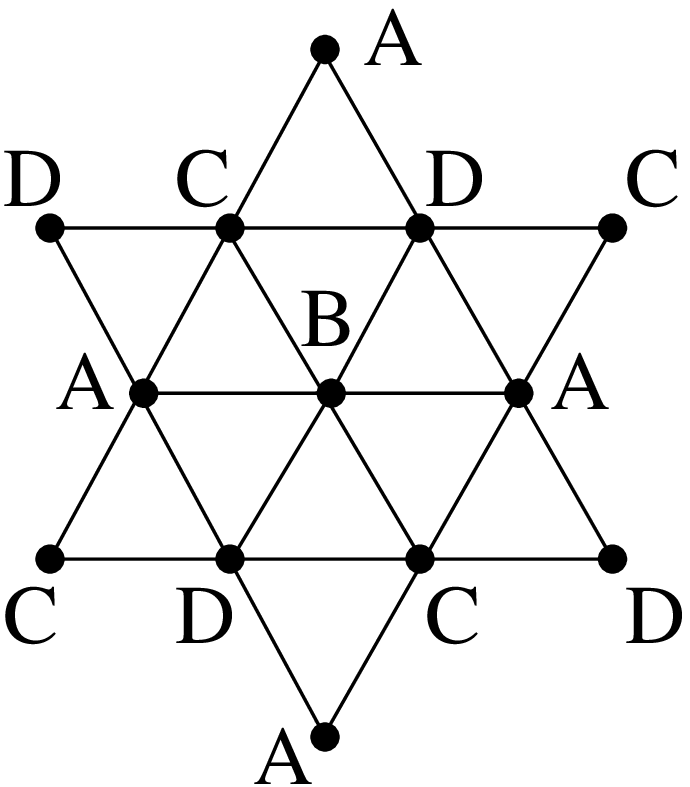}}
\quad
\subfigure[]{\includegraphics[height=2.7cm]{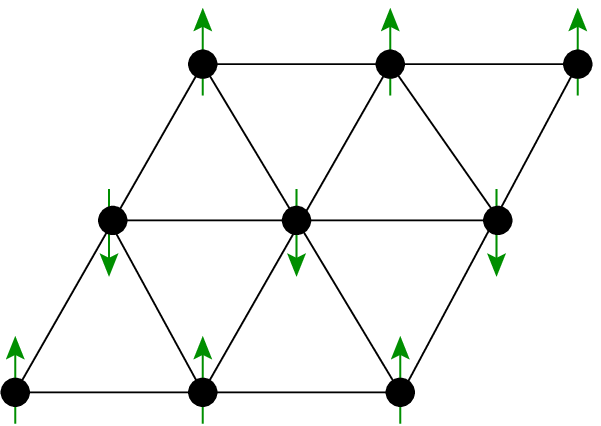}}
}
\caption{(Colour on-line) The $J_{1}$--$J_{2}$ model on the triangular
  lattice, showing (a) the bonds ($J_{1}
  \equiv$ -----~; $J_{2} \equiv \textcolor{blue}{- - -}$) and the
  Bravais lattice vectors $\hat{a}$ and $\hat{b}$; (b) the corresponding first Brillouin zone; (c) the
  120$^{\circ}$ N\'{e}el antiferromagnetic (AFM) state; (d) the
  infinitely degenerate family of classical 4-sublattice ground states
  on which the spins on the same sublattice are parallel to each other,
  with the sole constraint that the sum of four spins on different
  sublattices is zero; and (e) one of the three degenerate striped AFM
  states.  For the two states shown the arrows represent the
  directions of the spins located on lattice sites \textbullet.}
\label{model}
\end{center}
\end{figure*}  

\section{INTRODUCTION}
\label{introd_sec}
Much theoretical progress has been made in the search for exotic states in frustrated two-dimensional (2D)
quantum spin-lattice models that have no classical counterparts.  A special role is reserved for the 2D triangular lattice,
which is the simplest of the Archimedean lattices that exhibit
geometric frustration ({\it i.e.}, which are not bipartite) and the one with
the highest coordination number.  For the
triangular-lattice Heisenberg antiferromagnet (HAFM), comprising spins
(with spin quantum number $s$) interacting via a pure
nearest-neighbour (NN) Heisenberg potential with exchange coupling
$J_{1}>0$, the classical ($s \rightarrow \infty$) ground state (GS) is
a 3-sublattice N\'{e}el state with an angle of 120$^{\circ}$ between
the spins on different sublattices, thereby breaking the translational
symmetry of the lattice.  When Anderson first introduced
the concept of a resonating valence-bond (RVB) state
\cite{Anderson:1973_QSL,Fazekas:1972_QSL}, it was hypothesized that
the GS of the spin-$\frac{1}{2}$ HAFM on the triangular lattice with
NN interactions only, could be such an RVB state, the
first example of the quantum spin-liquid (QSL) states
\cite{Balents:2010_QSL}.

This conjecture was not supported by 
spin-wave theory (SWT), both at lowest (first) and second orders of
implementation, which predict that the system retains the 3-sublattice
120$^{\circ}$ N\'{e}el antiferromagnetic (AFM) long-range (LRO),
albeit with a reduction in the sublattice magnetization ({\it i.e.}, the
magnetic order parameter) of about 50\% for the $s=\frac{1}{2}$ case
from the classical ($s \rightarrow \infty$) value
\cite{Oguchi:1983_triang_SWT,Nishimori:1985_triang_SWT,Jolicoeur:1989_triang_SWT,Miuyaki:1992_triang_SWT,Chubukov:1994_triang_SWT,Chernyshev:2009_triang_SWT}.
Many other numerical studies, including variational calculations
\cite{Huse:1988_triang_VMC,Sindzingre:1994_triang_VMC}, and some based
on the exact diagonalization (ED) of small lattice clusters
\cite{Bernu:1992_triang_ED,Bernu:1994_triang_ED}, lent credence to the
SWT findings.  A variety of recent high-accuracy numerical studies,
including some based on the series expansion (SE) method
\cite{Zheng:2006_triang_ED}, the ED method
\cite{Richter:2004_triang_ED}, the Green's function Monte Carlo (GFMC)
method \cite{Capriotti:1999_trian}, the density-matrix renormalization
group (DMRG) method \cite{White:2007_triang_DMRG}, and the coupled
cluster method (CCM)
\cite{DJJFarnell:2014_archimedeanLatt,Li:2015_j1j2-triang}, all now
concur that the GS of the spin-$\frac{1}{2}$ triangular-lattice HAFM
with NN couplings only has the 3-sublattice 120$^{\circ}$ N\'{e}el
ordering, with a sublattice magnetization in the rather accurately determined
range of ($40 \pm 2)\%$ of the classical value.

To investigate a possible instability of the 120$^{\circ}$ N\'{e}el
phase against the formation of more exotic GS phases, it is natural to
add additional dynamic frustration by, for example, including an
additional Heisenberg interaction between next-nearest-neighbour (NNN)
pairs of spins with an exchange coupling $J_{2} \equiv \kappa J_{1}$.
The study of the resulting spin-$\frac{1}{2}$ $J_{1}$--$J_{2}$ model
on the triangular lattice is the main aim of the present paper,
particularly in the most interesting region of the parameters, $J_{1}
> 0$ and $0 \leq \kappa \leq 1$.

A number of recent studies
\cite{Li:2015_j1j2-triang,Manuel:1999_J1J2triang,Mishmash:2013_J1J2triang,Kaneko:2014_J1J2triang,Zhu:2015_j1j2-triang,Hu:2015_j1j2-triang}
have shown the existence of a magnetically disordered quantum state in
a range $\kappa^{c}_{1} < \kappa < \kappa^{c}_{2}$, beyond the range
$\kappa < \kappa^{c}_{1}$ for which the 120$^{\circ}$ N\'{e}el
ordering persists, and around the corresponding point $\kappa^{{\rm
    cl}}_{1} = \frac{1}{8}$ in the classical ($s \rightarrow \infty$)
version of the model at which the 120$^{\circ}$ N\'{e}el order melts.
Nevertheless, there remains disagreement over the values of the
quantum critical points (QCPs) $\kappa^{c}_{1}$ and $\kappa^{c}_{2}$,
and also the nature of this intermediate phase.  For example, a
Schwinger boson mean-field theory (SBMFT) study
\cite{Manuel:1999_J1J2triang} gives $\kappa^{c}_{1} \approx 0.12$ and
$\kappa^{c}_{2} \approx 0.19$.  By contrast, two different variational
Monte Carlo (VMC) studies give $\kappa^{c}_{1} \approx 0.05$ and
$\kappa^{c}_{2} \approx 0.18$ \cite{Mishmash:2013_J1J2triang}, and
$\kappa^{c}_{1} \approx 0.10(1)$ and $\kappa^{c}_{2} \approx 0.135(5)$
\cite{Kaneko:2014_J1J2triang}.  Both VMC studies favour the
intermediate state to be a gapless QSL, although of different types.
However, both the SMBFT and VMC techniques involve either
uncontrolled approximations and/or inbuilt bias towards certain forms
of order.

Two essentially unbiased {\it ab initio}
methods, with controlled approximations, that have been applied to the
model are the CCM \cite{Li:2015_j1j2-triang} and the DMRG
\cite{Zhu:2015_j1j2-triang,Hu:2015_j1j2-triang,Saadatmand:2015_j1j2-triang}.
Three recent studies give good agreement on the
QCPs.  The CCM study \cite{Li:2015_j1j2-triang} gives
$\kappa^{c}_{1} \approx 0.060(10)$ and $\kappa^{c}_{2} \approx
0.165(5)$, while two separate DMRG studies give $\kappa^{c}_{1}
\approx 0.06$ and $\kappa^{c}_{2} \approx 0.17$
\cite{Zhu:2015_j1j2-triang}, and $\kappa^{c}_{1} \approx 0.08$ and
$\kappa^{c}_{2} \approx 0.16$ \cite{Hu:2015_j1j2-triang}.  (In the third recent DMRG study \cite{Saadatmand:2015_j1j2-triang} the
model was studied on a three-leg cylinder, for which no intermediate
phase was found.)  The two DMRG studies
\cite{Zhu:2015_j1j2-triang,Hu:2015_j1j2-triang} both favour the
intermediate state to be a gapped QSL.  Since the nature of the
intermediate state has not yet been investigated within the CCM, which
method probably provides the most accurate values available of the
QCPs $\kappa^{c}_{1}$ and $\kappa^{c}_{2}$, our primary aim here is to
study it via its spin gap.  Our main finding will be that, in
sharp contrast to both recent DMRG studies
\cite{Zhu:2015_j1j2-triang,Hu:2015_j1j2-triang}, the intermediate
state is {\it gapless} over the entire regime $\kappa^{c}_{1} < \kappa
< \kappa^{c}_{2}$, just as was found in the two earlier VMC studies
\cite{Mishmash:2013_J1J2triang,Kaneko:2014_J1J2triang}.

\section{THE MODEL}
\label{model_sec}
We study the Hamiltonian
\begin{equation}
H = J_{1}\sum_{\langle i,j \rangle} \mathbf{s}_{i}\cdot\mathbf{s}_{j} + J_{2}\sum_{\langle\langle i,k \rangle\rangle} \mathbf{s}_{i}\cdot\mathbf{s}_{k}\,,
\label{eq1}
\end{equation}
where $\langle i,j \rangle$ and $\langle\langle i,k \rangle\rangle$
run over NN and NNN pairs of sites, respectively, counting each bond
once only.  Each site $i$ of the triangular lattice (with
lattice spacing $a$) carries a spin-$\frac{1}{2}$ particle with spin
operator ${\bf s}_{i} = (s^{x}_{i},s^{y}_{i},s^{z}_{i})$.  The lattice and the exchange bonds are
illustrated in fig.\ \ref{model}(a).
The first Brillouin zone for the triangular lattice is a hexagon of
side $4\pi/3a$, as shown in fig.\ \ref{model}(b).

Classically, the spin configurations can be described by a wave vector
$\mathbf{Q}$, such that the spin on site $i$ is given by

\begin{equation}
\mathbf{s}_{i}=s[\cos(\mathbf{Q}\cdot\mathbf{r}_{i})\hat{n}_{1}+\sin(\mathbf{Q}\cdot\mathbf{r}_{i})\hat{n}_{2}]\,,  \label{eq_classical-spins}
\end{equation}
where $\hat{n}_{1}$ and $\hat{n}_{2}$ are two orthogonal unit vectors
in spin space.  In the regime $J_{1}>0$, and $0 \leq \kappa \leq 1$,
where $\kappa \equiv J_{2}/J_{1}$, the corresponding classical ($s
\rightarrow \infty$) phase diagram has just one phase transition at
$\kappa^{{\rm cl}}_{1} = \frac{1}{8}$.  For $\kappa < \kappa^{{\rm
    cl}}_{1}$ the system has the 3-sublattice 120$^{\circ}$ N\'{e}el
ordering shown in fig.\ \ref{model}(c), described by either one of the
two wave vectors representing the two inequivalent corners $A$ and $B$
of the first Brillouin zone shown in fig.\ \ref{model}(b),
$\mathbf{Q}_{A} = \frac{4\pi}{a}(\frac{1}{3},0)$, $\mathbf{Q}_{B} =
\frac{2\pi}{a}(\frac{1}{3},\frac{1}{\sqrt{3}})$.  For $1 > \kappa >
\kappa^{c}_{1}$, the classical system has an infinitely degenerate
family (IDF) of 4-sublattice N\'{e}el GS phases illustrated in fig.\
\ref{model}(d), in which the sole constraint is $\mathbf{s}_{A} +
\mathbf{s}_{B} + \mathbf{s}_{C} + \mathbf{s}_{D} = 0$, where
$\mathbf{s}_{i}$ denotes the spin on each of the four sublattices $i =
A,B,C,D$, as shown.  The application of lowest-order [{\it i.e.}, to
$O(1/s)$] SWT has shown \cite{Joliceur:1990_J1J2-triang,AChubukov:1992_J1J2triang,SEKorshunov:1993_J1J2triang} that quantum fluctuations lift this accidental degeneracy in favour of the 2-sublattice striped states, one of which is shown in fig.\ \ref{model}(e), via the well-known order by disorder mechanism.  These striped states have ferromagnetic ordering along one principal direction of the triangular lattice and AFM ordering along the other two principal directions.  They are described by one of the three wave vectors representing the three inequivalent midpoints $K$, $L$, and $M$ of the edges of the hexagonal first Brillouin zone shown in fig.\ \ref{model}(b), $\mathbf{Q}_{K} = \frac{2\pi}{a}(0,\frac{1}{\sqrt{3}})$, $\mathbf{Q}_{L} = \frac{\pi}{a}(-1,\frac{1}{\sqrt{3}})$, and $\mathbf{Q}_{M} = \frac{\pi}{a}(1,\frac{1}{\sqrt{3}})$.

In our present study we will use both classical GS phases,
{\it viz.}, the 3-sublattice 120$^{\circ}$ N\'{e}el phase and the
2-sublattice striped state, as reference states, on top of
which the quantum many-body correlations for the $s=\frac{1}{2}$ model
will be systematically incorporated using the CCM.

\section{THE COUPLED CLUSTER METHOD}
\label{ccm_sec}
The CCM (see, {\it e.g.}, refs.\
\cite{Bishop:1991_TheorChimActa_QMBT,Bishop:1998_QMBT_coll,Fa:2004_QM-coll})
is one of the most versatile, most accurate and most powerful techniques of {\it ab
  initio} quantum many-body theory.  It has been applied to a
huge variety of fields in physics and chemistry, including many
strongly correlated and highly frustrated spin-lattice problems in
quantum magnetism (see, {\it e.g.}, refs.\
\cite{DJJFarnell:2014_archimedeanLatt,Li:2015_j1j2-triang,Fa:2004_QM-coll,Zeng:1998_SqLatt_TrianLatt,Kruger:2000_JJprime,SEKruger:2006_spinStiff,Bi:2008_EPL_J1J1primeJ2_s1,Li:2012_anisotropic_kagomeSq,Bishop:2014_honey_XY}).  Whereas many alternative methods require a finite-size
scaling analysis, one of the main strengths of the CCM is that it is
both size-consistent and size-extensive, and hence yields results from
the outset in the 2D bulk ($N \rightarrow
\infty$) limit.  Another important feature is that it
exactly preserves the important Hellmann-Feynman theorem at {\it all}
levels of approximate implementation.  The method
is utilized in practice at various levels of approximation, each specified by a truncation index $m =
1,2,3,\cdots$.  The {\it only} approximation made in using of the
CCM is then to extrapolate the sequences of values obtained for
physical observables to the exact $m \rightarrow \infty$ limit.

Every implementation of the CCM starts with the choice of a
suitable (normalized) reference (or model) state $|\Phi\rangle$, with
respect to which the quantum corrections present in the exact GS
$|\Psi\rangle$ may then be incorporated at the next step.  Suitable
choices for $|\Phi\rangle$ in the present case are the two
quasiclassical AFM states [{\it viz.}, the 3-sublattice 120$^{\circ}$
N\'{e}el state of fig.\ \ref{model}(c) and the 2-sublattice striped
state of fig.\ \ref{model}(e)].  The exact ket and bra GS wave functions $|\Psi\rangle$ and
$\langle\tilde{\Psi}|$, which solve the respective Schr\"{o}dinger
equations $H|\Psi\rangle = E|\Psi\rangle$ and $\langle\tilde{\Psi}|H =
E\langle\tilde{\Psi}|$, are chosen to have normalizations such that
$\langle\tilde{\Psi}|\Psi\rangle = \langle\Phi|\Psi\rangle =
\langle\Phi|\Phi\rangle = 1$.  In terms of the chosen model state the CCM now utilizes its distinct
exponential parametrizations, $|\Psi\rangle = e^{S}|\Phi\rangle$ and
$\langle\tilde{\Psi}| = \langle\Phi|\tilde{S}e^{-S}$.  The two GS
correlation operators are themselves formally decomposed as $S=\sum_{I
  \neq 0}{\cal S}_{I}C^{+}_{I}$ and $\tilde{S}=1+\sum_{I \neq
  0}\tilde{{\cal S}}_{I}C^{-}_{I}$, where we define $C^{+}_{0} \equiv
1$ to be the identity operator, and where the set index $I$ denotes a
complete set of single-particle configurations for all of the $N$
particles ({\it i.e.}, in our case, spins).

For spin-lattice applications each set index $I$ represents
a unique multispin-flip configuration with respect to 
$|\Phi\rangle$, such that the corresponding wave function for this
configuration of spins is $C^{+}_{I}|\Phi\rangle$.  We thus require
$|\Phi\rangle$ to be chosen as a fiducial (or cyclic)
vector ({\it i.e.}, in physical terms, a generalized vacuum state)
with respect to the complete set of mutually commuting many-body
creation operators $\{C^{+}_{I}\}$.  In other words, we require
$C^{+}_{I}$ and its destruction counterpart, $C^{-}_{I} \equiv
(C^{+}_{I})^{\dagger}$, to satisfy the conditions
$\langle\Phi|C^{+}_{I} = 0 = C^{-}_{I}|\Phi\rangle,\,\forall I \neq
0$.

It is useful to treat each lattice site $i$
as equivalent to all others, whatever the choice of 
$|\Phi\rangle$.  To that end we passively rotate each
spin $\mathbf{s}_{i}$ so that in its own local spin-coordinate frame
it points in the downward ({\it i.e.}, negative $z_{s}$) direction.
Henceforth our description of the spins is given in terms of
these locally defined spin-coordinate frames, in which all
independent-spin product model states take the universal form
$|\Phi\rangle = |\downarrow\downarrow\downarrow\cdots\downarrow\rangle$.
Similarly, $C^{+}_{I}$ may be written in the universal form $C^{+}_{I}
\equiv s^{+}_{k_{1}}s^{+}_{k_{2}}\cdots
s^{+}_{k_{n}};\,n=1,2,\cdots,2sN$, as a product of single-spin raising
operators, $s^{+}_{k} \equiv s^{x}_{k} + is^{y}_{k}$.  The set index
$I \equiv \{k_{1},k_{2},\cdots k_{n}; n=1,2,\cdots,2sN\}$ thus becomes a set of (possibly repeated) lattice site indices.  For
the present case, with $s=\frac{1}{2}$, each site index included in set index $I$ may appear no more than once.  Once the local spin
coordinates have been chosen for a given $|\Phi\rangle$,
the Hamiltonian $H$ needs to be re-expressed in terms of them.

The GS CCM $c$-number correlation coefficients $\{{{\cal
    S}}_{I},\tilde{{\cal S}}_{I}\}$ are now calculated by requiring
that the GS energy expectation functional, $\bar{H} = \bar{H}[{\cal
  S}_{I},\tilde{{\cal S}}_{I}] \equiv \langle\Phi|\tilde{S}e^{-S}He^{S}|\Phi\rangle$, be minimized with respect to each of
them separately.  This yields the coupled set of nonlinear equations
$\langle\Phi|C^{-}_{I}e^{-S}He^{S}|\Phi\rangle = 0;\,\forall I \neq
0$, for the coefficients $\{{\cal S}_{I}\}$, and the coupled set of
linear generalized eigenvalue equations
$\langle\Phi|\tilde{S}[e^{-S}He^{S},C^{+}_{I}]|\Phi\rangle = 0$; or
equivalently, $\langle\Phi|\tilde{S}(e^{-S}He^{S}-E)C^{+}_{I}|\Phi\rangle =
0,\,\forall I \neq 0$, for the coefficients $\{\tilde{{\cal
    S}}_{I}\}$, once the coefficients $\{{\cal S}_{I}\}$ are solved
for.  Having solved for $\{{{\cal S}}_{I},\tilde{{\cal S}}_{I}\}$, one
may then, for example, calculate all GS properties such as the energy
$E=\langle\Phi|e^{-S}He^{S}|\Phi\rangle$ and the magnetic order
parameter ({\it i.e.}, the average local on-site magnetization) $M \equiv
-\frac{1}{N}\langle\Phi|\tilde{S}\sum^{N}_{k=1}e^{-S}s^{z}_{k}e^{S}|\Phi\rangle$.

Excited-state (ES) wave functions are parametrized within the CCM
as $|\Psi_{e}\rangle = X^{e}e^{S}|\Phi\rangle$, in terms of a linear
excitation operator $X^{e}=\sum_{I \neq 0}{\cal X}^{e}_{I}C^{+}_{I}$.  By combining the GS Schr\"{o}dinger equation with its ES
counterpart, $H|\Psi_{e}\rangle = E_{e}|\Psi_{e}\rangle$, and
using that $[X^{e},S] = 0$, we find
$e^{-S}[H,X^{e}]e^{S}|\Phi\rangle = \Delta_{e}X^{e}|\Phi\rangle$,
where $\Delta_{e} \equiv (E_{e}-E)$ is the excitation energy.  By
taking the overlap with the state $\langle\Phi|C^{-}_{I}$ we find
$\langle\Phi|C^{-}_{I}[e^{-S}He^{S},X^{e}]|\Phi\rangle =
\Delta_{e}{\cal X}^{e}|\Phi\rangle;\, \forall I \neq 0$, where we have used
that the states labelled by the set indices $I$ are orthonormalized,
$\langle\Phi|C^{-}_{I}C^{+}_{J}|\Phi\rangle = \delta(I,J)$.  These generalized
eigenvalue equations may then be solved for the set $\{{\cal X}^{e}_{I}\}$ and
$\Delta$.

\begin{figure*}[t]
\begin{center}
\mbox{
\subfigure[]{\includegraphics[angle=270,width=7.5cm]{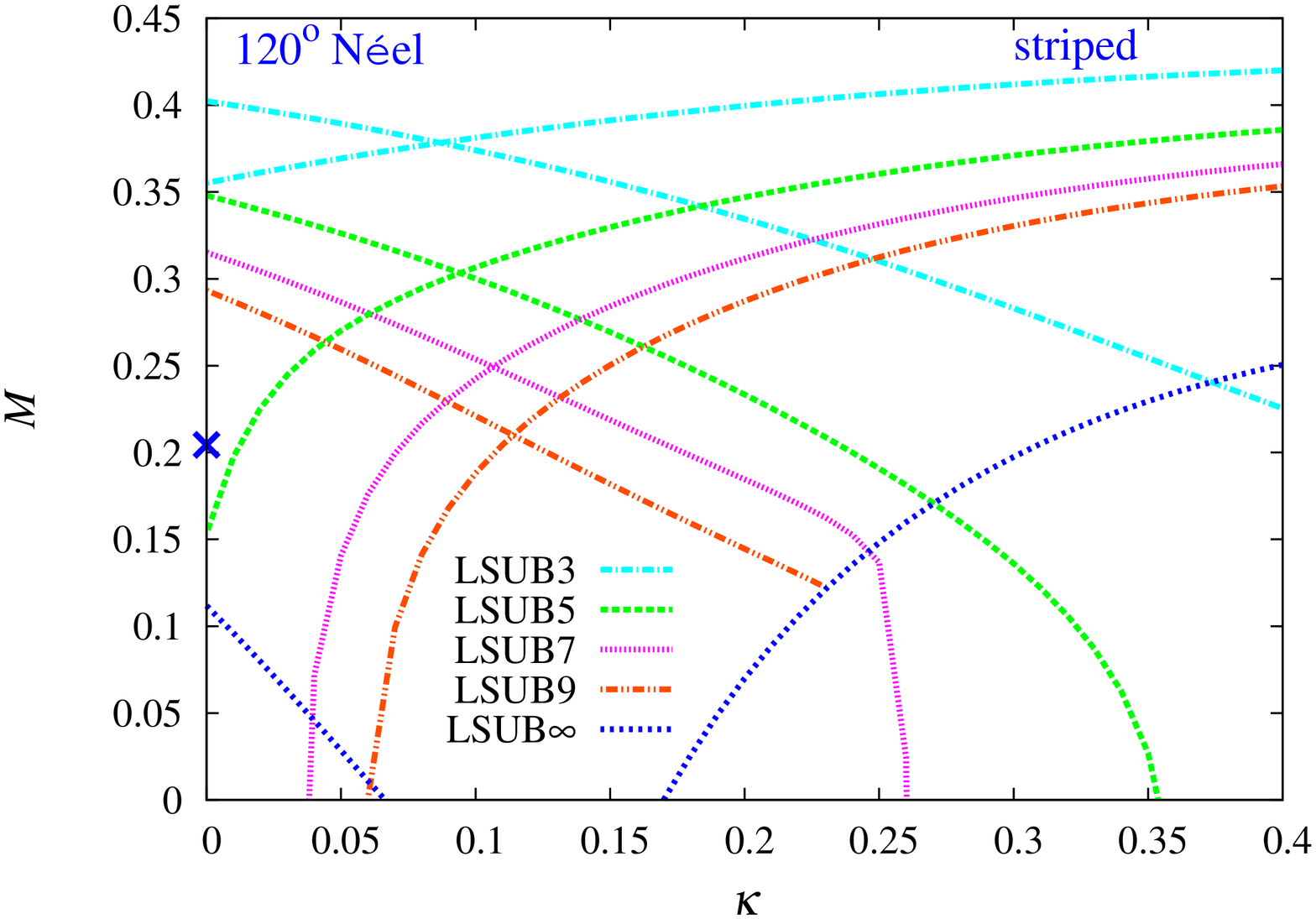}}
\quad
\subfigure[]{\includegraphics[angle=270,width=7.5cm]{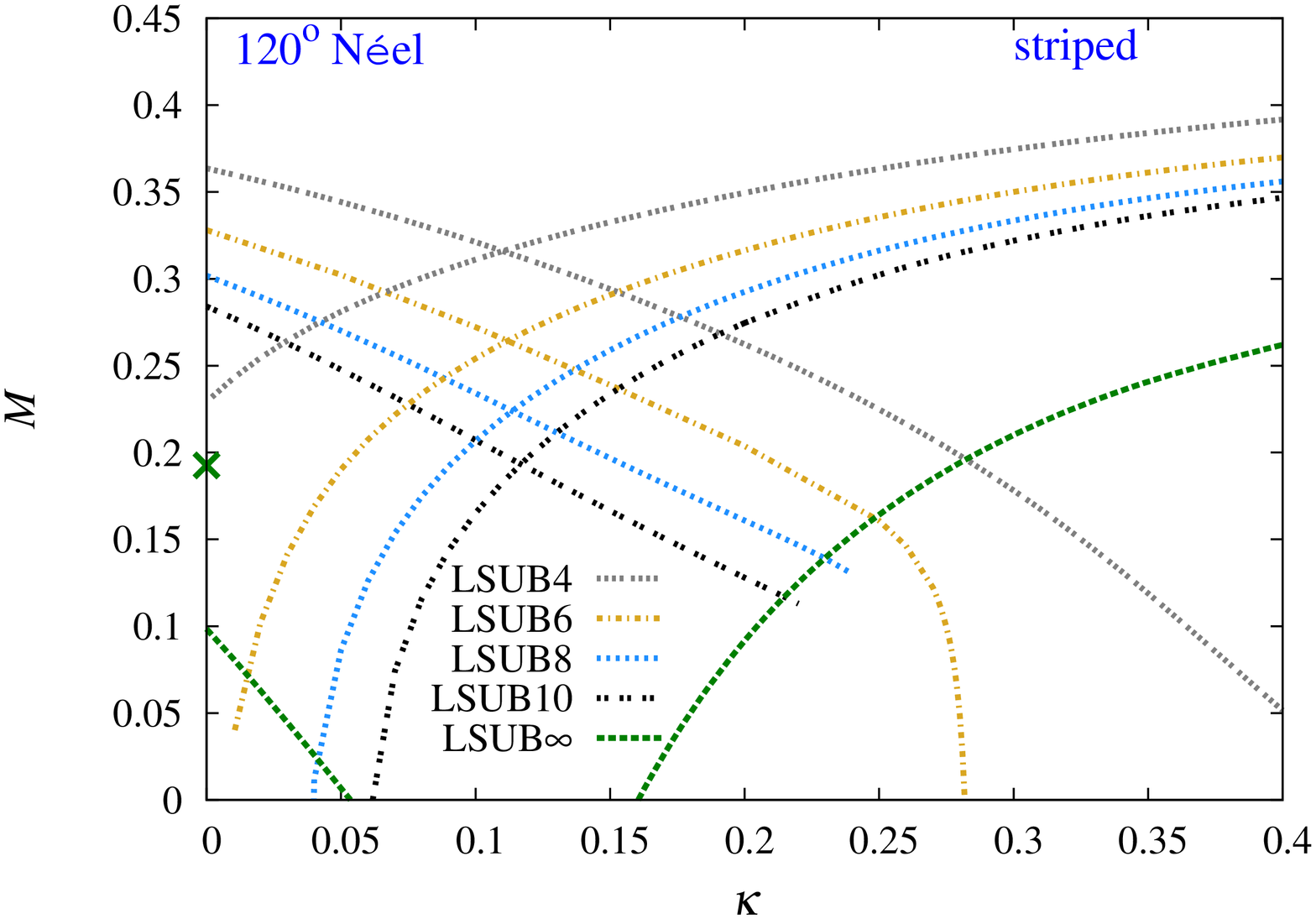}}
}
\caption{(Colour on-line) CCM results for the GS magnetic order
  parameter $M$ versus the frustration parameter $\kappa \equiv
  J_{2}/J_{1}$, for the spin-$\frac{1}{2}$ $J_{1}$--$J_{2}$ model on
  the triangular lattice (with $J_{1} > 0$).  The left and right sets
  of curves in each panel are based on the 120$^{\circ}$ N\'{e}el and
  striped AFM states, respectively, as CCM model states.  The LSUB$m$
  results are shown: (a) with $m=\{3,5,7,9\}$; and (b) with
  $m=\{4,6,8,10\}$.  In both cases the extrapolated LSUB$\infty$
  results using the respective data set in eq.\
  (\ref{M_extrapo_frustrated}) are shown.  We also show with cross
  ($\times$) symbols the corresponding extrapolated values based on
  the 120$^{\circ}$ N\'{e}el state using eq.\
  (\ref{M_extrapo_standard}) for the case $\kappa=0$ ({\it i.e.}, the
  pure triangular-lattice HAFM with NN interactions only).}
\label{M}
\end{center}
\end{figure*}

No approximations have yet been made.  One might expect that
truncations are needed in the evaluation of the exponential functions
$e^{\pm S}$ in the CCM parametrizations.
However, we note that these always appear only in the form of the
similarity transform $e^{-S}He^{S}$ of the Hamiltonian in all of the
GS or ES equations we need to solve.  This similarity transform may be
expanded in terms of the well-known nested commutator series.  Another
key feature of the CCM is that this otherwise infinite series actually
now terminates exactly at the double commutator term.  This is due to
the basic SU(2) commutation relations and the fact that all of the
terms in the expansion $\sum_{I \neq 0}{\cal S}_{I}C^{+}_{I}$ for $S$
commute with one another and are simple products of single-spin
raising operators.  Furthermore, all terms in the expansion for
$e^{-S}He^{S}$ are thus linked.  Hence, the Goldstone linked cluster
theorem and the consequent size-extensivity of the method are exactly preserved even if the expansion for $S$ is
truncated.

Thus, the {\it only} approximation made in practical implementations
of the CCM is to restrict the set of indices $\{I\}$
retained in the corresponding expansions for the operators $\{S,
\tilde{S}\}$.  As in our previous work on this model
\cite{Li:2015_j1j2-triang}, and in many other applications of the CCM
to quantum magnets, we use here the well-tested localized
(lattice-animal-based subsystem) LSUB$m$ scheme.  At the $m$th level of approximation all multispin-flip
configurations $\{I\}$ are retained in the expansions for $S$ and
$\tilde{S}$ that are constrained to no more than $m$ contiguous
sites.  A cluster configuration is contiguous if every site in it is
NN to at least one other.  By utilizing the space- and point-group
symmetries of the model, together with any conservation laws that
apply to both the Hamiltonian and the model state under study, we may
reduce the terms retained to a minimal number $N_{f} = N_{f}(m)$ of
fundamental LSUB$m$ configurations.  Nevertheless, $N_{f}(m)$
increases rapidly as the truncation index $m$ is increased, and one
soon needs to use massive parallelization and supercomputing resources
for the higher-order calculations
\cite{Zeng:1998_SqLatt_TrianLatt,ccm_code}.

For the ES calculation of the (magnon, triplet) spin gap $\Delta$ the
choice of cluster configuration $\{I\}$ retained in the expansion for
the excitation operator $X^{e}$ is different from those for the GS
correlation operators $S$ and $\tilde{S}$, and different too for each model state, since we now restrict
attention to those that change the $z$ component of total spin,
$S^{z}$, by one unit (in the local spin-coordinate frames), rather than to those with $S^{z}=0$ for the GS
calculation.  However, to ensure comparable accuracy for both the GS
and ES calculations, we use the LSUB$m$ scheme in each case.  The
number $N_{f}(m)$, for a given value of $m$, is appreciably higher for
the ES case than for the corresponding GS case using the same model
state.  The corresponding GS CCM LSUB$m$ equations have been solved
for both model states for $m \leq 10$.  Whereas we can also solve the
ES LSUB$m$ equations for $m \leq 10$ for the striped state, the number
$N_{f}(10)$ for the ES using the 120$^{\circ}$ N\'{e}el state is now
sufficiently large that we are restricted in practice to $m \leq 9$ in this case.

\begin{figure*}[t]
\begin{center}
\mbox{
\subfigure[]{\includegraphics[angle=270,width=7.5cm]{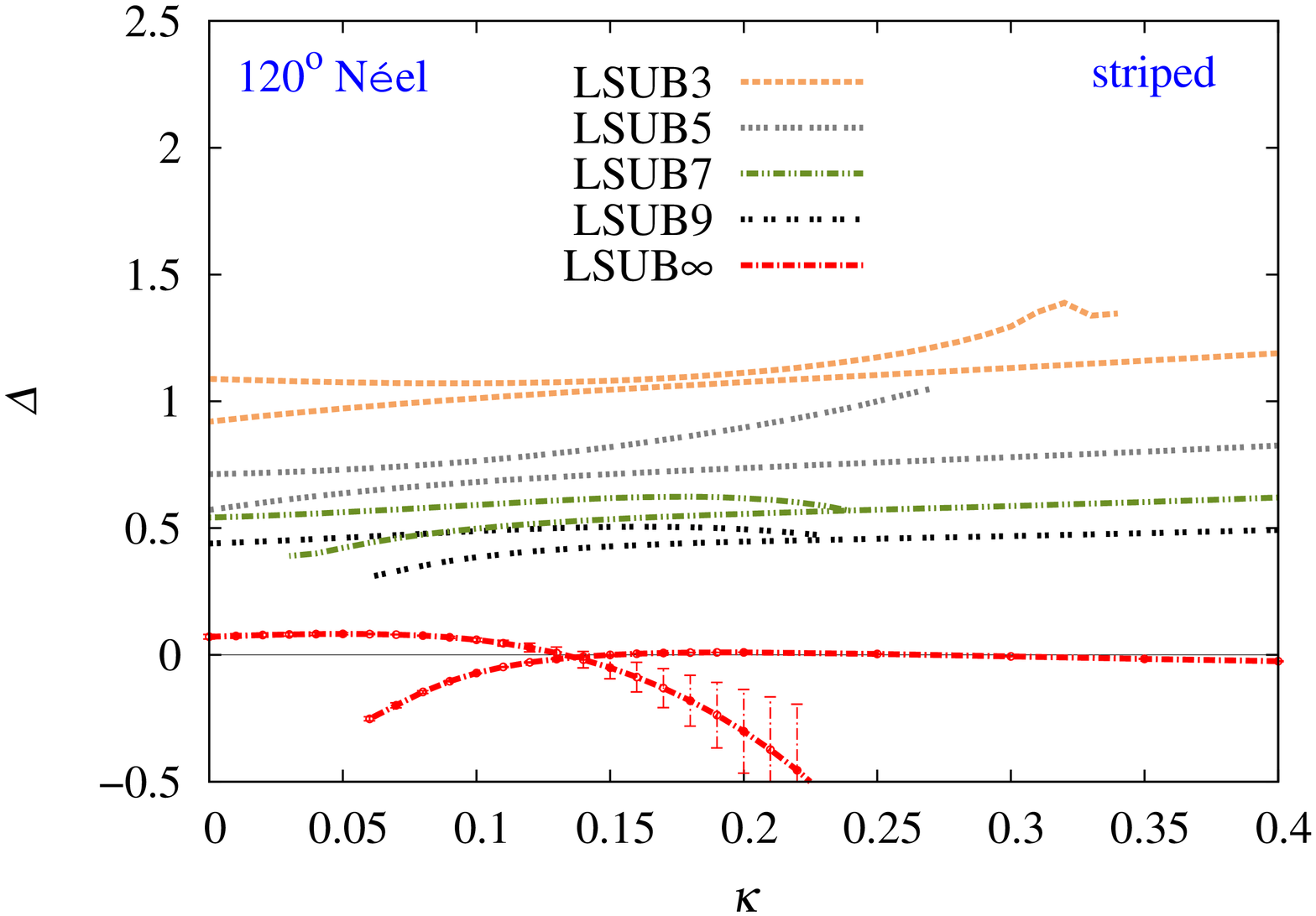}}
\quad
\subfigure[]{\includegraphics[angle=270,width=7.5cm]{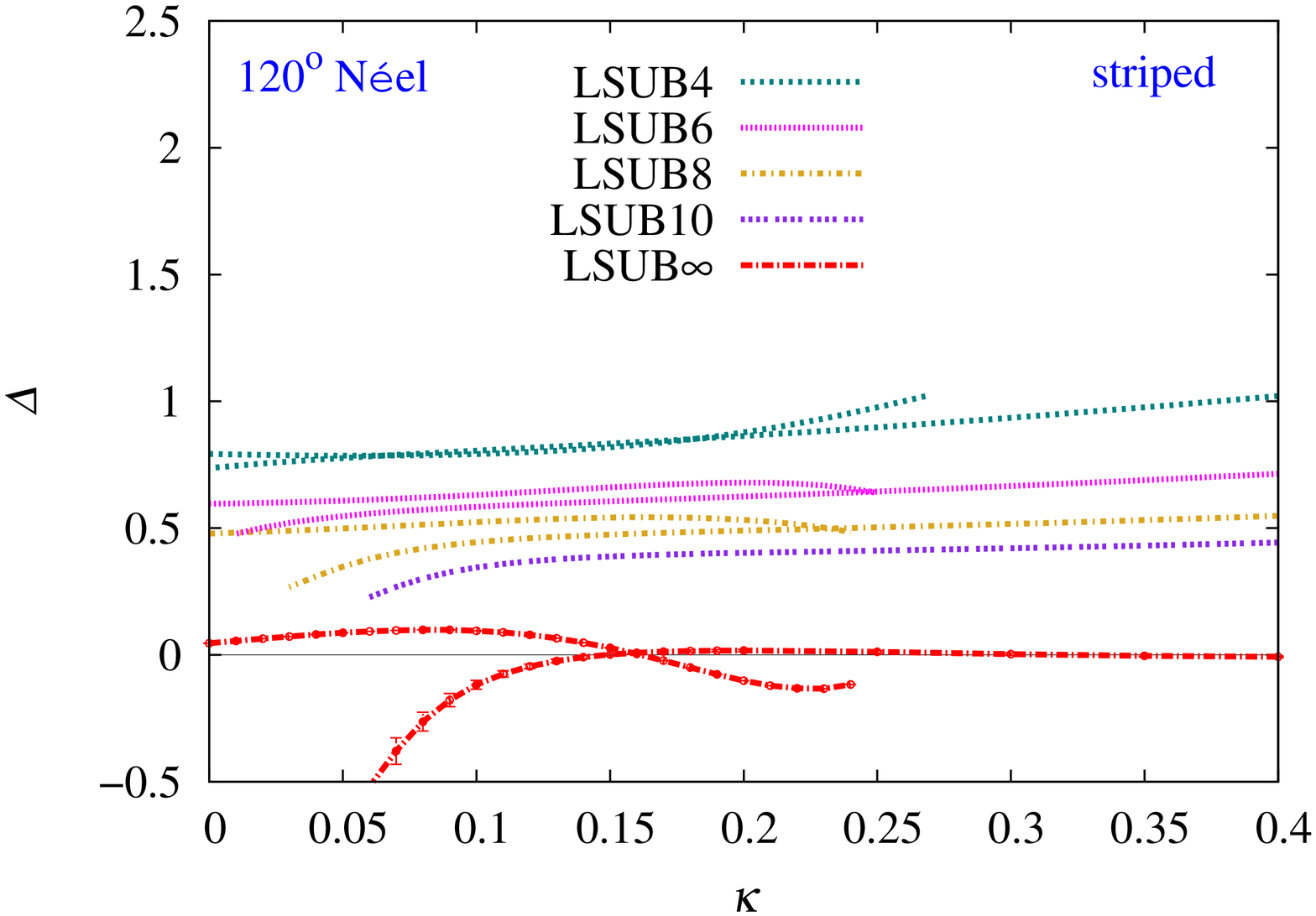}}
}
\caption{(Colour on-line) CCM results for the spin gap $\Delta$ versus
  the frustration parameter $\kappa \equiv J_{2}/J_{1}$, for the
  spin-$\frac{1}{2}$ $J_{1}$--$J_{2}$ model on the triangular lattice
  (with $J_{1}=1$).  The left and right sets of curves in each panel
  are based on the 120$^{\circ}$ N\'{e}el and striped AFM states,
  respectively, as CCM model states.  The LSUB$m$ results are shown:
  (a) with $m=\{3,5,7,9\}$; and (b) with $m=\{4,6,8\}$ for the
  120$^{\circ}$ N\'{e}el state and $m=\{4,6,8,10\}$ for the striped
  state.  In both cases the extrapolated LSUB$\infty$ results using
  the respective data sets in eq.\ (\ref{Eq_spin_gap}) are shown,
  together with error bars associated with the fits using
  four data points.}
\label{E_gap}
\end{center}
\end{figure*}

Clearly, the CCM LSUB$m$ approximations become exact, by construction,
in the $m \rightarrow \infty$ limit.  Thus, although the CCM obviates
the need for any finite-size scaling, we do need to extrapolate our
LSUB$m$ results to the $m \rightarrow \infty$ limit as our last step
and as our sole approximation.  There exist very well-tested accurate
extrapolation rules for such GS quantities as the energy per spin,
$E/N$, and the magnetic order parameter, $M$, as we have described and
used in our previous paper \cite{Li:2015_j1j2-triang} on this model.
For example, as we discussed there, for unfrustrated models or ones
with relatively minor amounts of frustration we use
\begin{equation}
M(m) = b_{0} + b_{1}m^{-1} + b_{2}m^{-2}\,,   \label{M_extrapo_standard}
\end{equation}
whereas for highly frustrated systems, especially ones close to a QCP, we use
\begin{equation}
M(m) = b_{0} + b_{1}m^{-1/2} + b_{2}m^{-3/2}\,.   \label{M_extrapo_frustrated}
\end{equation}
For the spin gap we use the extrapolation scheme \cite{Kruger:2000_JJprime,Richter:2015_ccm_J1J2sq_spinGap}
\begin{equation}
\Delta(m) = d_{0}+d_{1}m^{-1}+d_{2}m^{-2}\,.   \label{Eq_spin_gap}
\end{equation}
Of course, for the GS expectation value of any parameter $P$, we may always fit our LSUB$m$ data to an asymptotic form $P(m)=p_{0}+p_{1}m^{-\nu}$, in order to extract the leading exponent $\nu$.  For our $\Delta(m)$ results, for example, we find $\nu \approx 1$ for all values of $\kappa$, which hence justifies the use of Eq.\ (\ref{Eq_spin_gap}).
Whenever possible we fit each of the extrapolation schemes of eqs.\
(\ref{M_extrapo_standard})--(\ref{Eq_spin_gap}), all of which contain
three fitting parameters, to LSUB$m$ results with at least four
different values of $m$.  Since the triangle is the basic structural
element of the lattice,
the lowest-order results with $m < 3$ are excluded from the
extrapolations.

\section{RESULTS}
\label{results_sec}
We show in fig.\ \ref{M} our results for the magnetic order parameter
$M$ for the model, using both the 120$^{\circ}$ N\'{e}el and striped
AFM states as CCM model states.
Since the LSUB$m$ results show a marked even-odd staggering effect in
the truncation index $m$, we display the odd and even values
separately in figs.\ \ref{M}(a) and \ref{M}(b), together with the
associated LSUB$\infty$ extrapolations using eq.\
(\ref{M_extrapo_frustrated}), which is appropriate in the highly
frustrated regions near the QCPs, together with the value using eq.\
(\ref{M_extrapo_standard}) for the unfrustrated case $\kappa = 0$ only.
The QCPs ({\it viz.}, the points $\kappa$ at which the extrapolated
LSUB$\infty$ values for $M$ vanish) show great consistency between
those based on even $m$ values and those based on odd $m$ values.  A
careful analysis of our errors gives our best estimates for the two
QCPs as $\kappa^{c}_{1}=0.060 \pm 0.010$ and $\kappa^{c}_{2}=0.165 \pm
0.005$.
We note, parenthetically, that the smooth shape of the $M(\kappa)$ curves in fig.\ \ref{M} near both QCPs lends support to both transitions being continuous.  However, since the extrapolations near any critical point are very sensitive, we cannot entirely rule out the possibility of either transition being first-order.

An interesting feature of our LSUB$m$ results displayed in fig.\
\ref{M} is that we find solutions based on a given model state over a
region that extends beyond the associated exact (LSUB$\infty$) QCP,
typically out to some associated termination point $\kappa^{t}(m)$,
beyond which, at a given LSUB$m$ level of approximation, no real
solution exists.  Such termination points have been found in many
other applications and are very well understood.  They have been shown
\cite{Fa:2004_QM-coll} to be direct manifestations of the associated
QCP (at $\kappa^{c},$ say) at which the order of the model state melts
in the physical system.  Typically the extent
$|\kappa^{t}(m)-\kappa^{c}|$ of the region of unphysical intrusion of
the solution with the order properties of the CCM model state into the
neighbouring phase (beyond the QCP at $\kappa^{c}$) in which this form
of order vanishes, decreases as the truncation index $m$ is increased,
and vanishes in the exact limit $m \rightarrow \infty$.  It is clear from fig.\ \ref{M} that these regions of unphysical
intrusion of the CCM LSUB$m$ solutions into the regime $\kappa^{c}_{1}
< \kappa < \kappa^{c}_{2}$ of the intermediate phase cover essentially
the whole intermediate region for all values $m \leq 10$ on both the
neighbouring 120$^{\circ}$ N\'{e}el and striped quasiclassical AFM
sides.  For that reason we are now well placed to investigate the
intermediate phase regime using both quasiclassical model states.

Thus, in figs.\ \ref{E_gap}(a) and \ref{E_gap}(b) we show our
corresponding LSUB$m$ results for the spin gap $\Delta$, together with the extrapolated LSUB$\infty$ values using eq.\
(\ref{Eq_spin_gap}).  
Goldstone's theorem implies that any state
breaking spin-rotational symmetry, including all states with
magnetic LRO, must have a vanishing gap.  We see from figs.\
\ref{E_gap}(a) and \ref{E_gap}(b) that our extrapolated LSUB$\infty$
results for $\Delta$ are wholly consistent with $\Delta=0$ for the
striped state, $\kappa > \kappa^{c}_{2} \approx 0.165$.  The results
on the (computationally more challenging) 120$^{\circ}$
N\'{e}el-ordered side, $\kappa < \kappa^{c}_{1} \approx 0.06$, are
slightly less accurate, but still compatible with Goldstone's theorem,
within small errors, presumably associated with extrapolations in
this region being less accurate.  

Most interestingly, the results in fig.\ \ref{E_gap} based on both
quasiclassical AFM states show no tendency at all for the spin gap
$\Delta$ to increase in the regions of the intermediate phase
accessible by the unphysical intrusion effect.  Of course, the spin gap, by definition, satisfies $\Delta \geq 0$.  The regions where $\Delta < 0$ in fig.\ \ref{E_gap} are clearly wholly unphysical therefore.  However, our results are invalid in the regions beyond where the curves based on each model state cross, since the value of $\Delta$ is, by definition, unique.  Beyond the crossing point we are simply too far into the unphysical regime, and we see from fig.\ \ref{E_gap} that this is reflected in the large error bars there.  We note too that our largest error for $\Delta$ on the 120$^{\circ}$ N\'{e}el ordered side (where $\Delta = 0$ is exact) is $\Delta \lesssim 0.1$, which is compatible with the values we obtain for $\Delta$ in the intermediate regime.  The extrapolated
values of the spin gap are thus compatible with the spin gap vanishing
essentially everywhere within the intermediate phase region,
$\kappa^{c}_{1} < \kappa < \kappa^{c}_{2}$.  For comparison, we note that the estimate obtained by a recent DMRG study \cite{Hu:2015_j1j2-triang} for the {\it bulk} triplet gap is $\Delta \approx 0.3 \pm 0.1$ at $\kappa = 0.1$, which is clearly at odds with our results.

\section{SUMMARY AND DISCUSSION}
Our primary finding is that we find no evidence at all for any part of
the intermediate phase to be a gapped state.  We may compare our study
with similar recent CCM spin-gap studies of both the
spin-$\frac{1}{2}$ $J_{1}$--$J_{2}$ model on the square lattice
\cite{Richter:2015_ccm_J1J2sq_spinGap} and the spin-$\frac{1}{2}$
$J_{1}$--$J_{2}$--$J_{3}$ model on the honeycomb lattice (with
$J_{3}=J_{2}$) \cite{Bishop:2015_honey_low-E-param}.  For the former
model $\Delta$ was again found to be zero over that part of the
equivalent intermediate phase region accessible by the unphysical
intrusion mechanism, which was not the whole region in this case.  By
contrast, for the latter model, strong evidence was found that $\Delta
\neq 0$ over the similarly accessible part of its equivalent
intermediate phase, which finding provides strong support for that
phase having valence-bond crystalline (VBC) order, as is now broadly
agreed.

Finally, in contrast to the recent DMRG studies
\cite{Zhu:2015_j1j2-triang,Hu:2015_j1j2-triang} that find the
intermediate phase of the spin-$\frac{1}{2}$ $J_{1}$--$J_{2}$ model on
the triangular lattice to be a gapped QSL, our CCM results indicate
that the intermediate phase is gapless.  While our results clearly
preclude the intermediate phase in this model to have any form of VBC
order, since any such state is gapped, they point towards a gapless
QSL as the most likely candidate.

\acknowledgments We thank the University of Minnesota Supercomputing
Institute for the grant of supercomputing facilities.  One of us (RFB) acknowledges the Leverhulme Trust for the award of an Emeritus Fellowship (EM-2015-07).

\bibliographystyle{eplbib}
\bibliography{bib_general}

\end{document}